\documentclass[twocolumn,secnumarabic,amssymb, nobibnotes, aps]{revtex4-1}

\usepackage{graphicx}
\usepackage{dcolumn}
\usepackage{bm}

\usepackage{color}
\usepackage{braket}

\begin{document}

\begin{abstract}
The electronic properties of graphene on top of ferroelectric HfO$_2$ substrates in orthorhombic phase with space group Pca2$_1$ are investigated using density functional theory calculations. The space group Pca2$_1$ was recently identified as one of the two potential candidates for ferroelectricity in hafnia, with the polarization direction oriented along [001] direction. Our results indicate the appearance of sizable energy gaps in graphene induced by the HfO$_2$ substrate, as a consequence of orbital hybridization and locally deformed graphene structure. The gap sizes depend on the type of the HfO$_2$ terminations interacting with graphene, showing larger gaps for oxygen terminated slabs compared to hafnium terminated ones. These observations may prove to be highly significant for the development of graphene based field effect transistors using high-k dielectrics.   
\end{abstract}

\title{Graphene bandgap induced by ferroelectric $Pca2_1$ HfO$_2$ substrate:\\ a first-principles study}

\author{George Alexandru Nemnes$^{1,2,*}$, Daniela Dragoman$^{1,3}$ and Mircea Dragoman$^4$}
\affiliation{$^1$University of Bucharest, Faculty of Physics, MDEO Research Center, 077125 Magurele-Ilfov, Romania}%
\email{E-mail: nemnes@solid.fizica.unibuc.ro}
\affiliation{$^2$Horia Hulubei National Institute for Physics and Nuclear Engineering, 077126 Magurele-Ilfov, Romania}%
\affiliation{$^3$Academy of Romanian Scientists, Splaiul Independentei 54, Bucharest 050094, Romania}%
\affiliation{$^4$National Research and Development Institute in Microtechnology, Str. Erou Iancu Nicolae 126A, Bucharest 077190, Romania}%

\maketitle

\section{Introduction}

In the search for high-$k$ dielectrics, the hafnium oxide (HfO$_2$) is an already established option to replace SiO$_2$, owing its high compatibility with silicon based CMOS technology. Besides nanoelectronic devices, HfO$_2$ is currently employed in a wide range of applications, such as optical coatings \cite{doi:10.1116/1.1382879},  passive radiative cooling \cite{Raman2014}, resistive-switching memories \cite{doi:10.1063/1.3567915} and passivation of silicon surfaces in solar cells \cite{doi:10.1063/1.4973988}. More recently, the discovery of ferroelectricity enriches the body of possible applications, as ferroelectric hafnium oxide (f-HfO$_2$) may be an alternative for replacing ternary perovskite materials in memory devices.

Ferroelectricity was evidenced in hafnium oxide thin films doped with Si \cite{doi:10.1063/1.3634052}, YO$_{1.5}$ \cite{doi:10.1063/1.4927450} or Y \cite{doi:10.1063/1.5020688} and in alloyed Hf$_x$Zr$_{1-x}$O$_y$ thin films \cite{doi:10.1021/nl302049k,doi:10.1021/acsami.5b11653}, with the most robust effect found for equal Hf and Zr fractions, i.e. Hf$_{0.5}$Zr$_{0.5}$O$_2$. Using group theory principles and density functional theory (DFT) calculations two orthorhombic phases with space groups $Pca2_1$ and $Pmn2_1$ were identified \cite{PhysRevB.90.064111}, which are non-centrosymmetric and thus compatible with ferroelectricity. Subsequently, combining aberration corrected high-angle annular dark-field scanning transmission electron microscopy and nanoscale electron diffraction methods, the structure of $Pca2_1$ f-HfO$_2$ was evidenced \cite{doi:10.1063/1.4919135}. Currently, the orthorhombic $Pca2_1$ structural phase of HfO$_2$ is thought to be primary cause for ferroelectricity \cite{doi:10.1002/admi.201701258}, although the $Pmn2_1$ phase with a slightly higher free-energy cannot be ruled out \cite{PhysRevB.90.064111}.

Down-scaling field-effect transistors requires high-k materials for gate dielectrics. In addition, graphene is a strong candidate for the new generations of high mobility FETs. However, inducing a gap in graphene is problematic, as the influence of a proper substrate inducing the symmetry breaking between A and B sublattices, such as hexagonal boron nitride (h-BN) \cite{PhysRevB.76.073103} or silicon carbide (SiC) \cite{Zhou2007}, may also affect the transport properties of graphene. Several oxides, amongst which one can mention Al$_2$O$_3$, Pb[Zr$_x$Ti$_{1-x}$]O$_3$  and HfO$_2$, have been considered for FET applications \cite{PhysRevLett.102.136808,doi:10.1116/1.3693416,KIM201385}. 
In FETs having ferroelectric substrates the mobility reaches 70000 cm$^2$ V$^{-1}$ s$^{-1}$ at room temperature due to its low roughness and due to ferroelectric built-in electric fields, beyond the state-of-art classical semiconductors, like A$^{\rm III}$-B$^{\rm V}$ compounds. Depositing high-quality HfO$_2$ on graphene indicated a reasonable mobility \cite{PhysRevLett.105.126601}, reaching values up to 20000 cm$^2$ V$^{-1}$ s$^{-1}$, but still one order of magnitude lower than in pristine graphene. The idea of using graphene on HfO$_2$ ferroelectric substrate was experimentally demonstrated in the case of memories with enhanced electrical properties \cite{Dragoman_2018}.

Understanding the processes leading to the mobility reduction and, in general, the transport properties of graphene based FETs requires an in-depth analysis of the interaction between graphene and the substrate. So far theoretical studies were focused on graphene on top of paraelectric forms of hafnia.  
Using DFT calculations the energetics and electronic structure of graphene adsorbed on cubic HfO$_2$ were investigated \cite{PhysRevB.83.153413,CHIU2014583,doi:10.1063/1.4961112}. The influence of the HfO$_2$ substrate on the transport properties was also analyzed, showing that the different terminations can significantly affect the transmission functions \cite{doi:10.1063/1.4961112}. In this context, a possible approach for mitigating these effects was indicated.

Here we investigate the electronic structure of graphene on top of ferroelectric HfO$_2$ in orthorhombic phase with space group $Pca2_1$, using first-principles calculations. We focus on the influence of the two possible terminations of the hafnia substrate oriented along [001] direction, which is also the polarization direction. Our analysis indicates the appearance of energy gaps in the graphene band structure, with typical sizes of $\sim$0.25 eV or larger, which depend on the nature of the surface terminations, hafnium or oxygen. The observed energy gaps may have a significant impact in the design of graphene based FETs, enabling the gate control over the source-drain current and allowing for significant ON/OFF current ratios.

\section{Structures and computational methods}

Although bulk HfO$_2$ occurs in natural form in monoclinic $P2_1/c$ phase, by increasing the temperature it transforms to the tetragonal $P4_2/nmc$ phase and subsequently to the cubic $Fm\bar{3}m$ phase \cite{Wang1992}. Ferroelectricity was observed in the orthorhombic hafnia which is typically stabilized by doping (e.g. Al, Y, La) \cite{doi:10.1063/1.3634052,doi:10.1116/1.4873323,doi:10.1063/1.5021746} or high pressure \cite{LIU1980331,PhysRevB.68.054106,Al_Khatatbeh_2018}. From the four proposed orthorhombic phases, $Pca2_1$, $Pmn2_1$, $Pbca$ and $Pbcm$, only the first two are non-centrosymmetric and can sustain ferroelectricity \cite{doi:10.1063/1.4919135}. Since the $Pca2_1$ phase was experimentally identified, in contrast to $Pmn2_1$ phase, we selected it for the subsequent investigation.

\begin{figure}[t]
\centering
\includegraphics[width=0.45\linewidth]{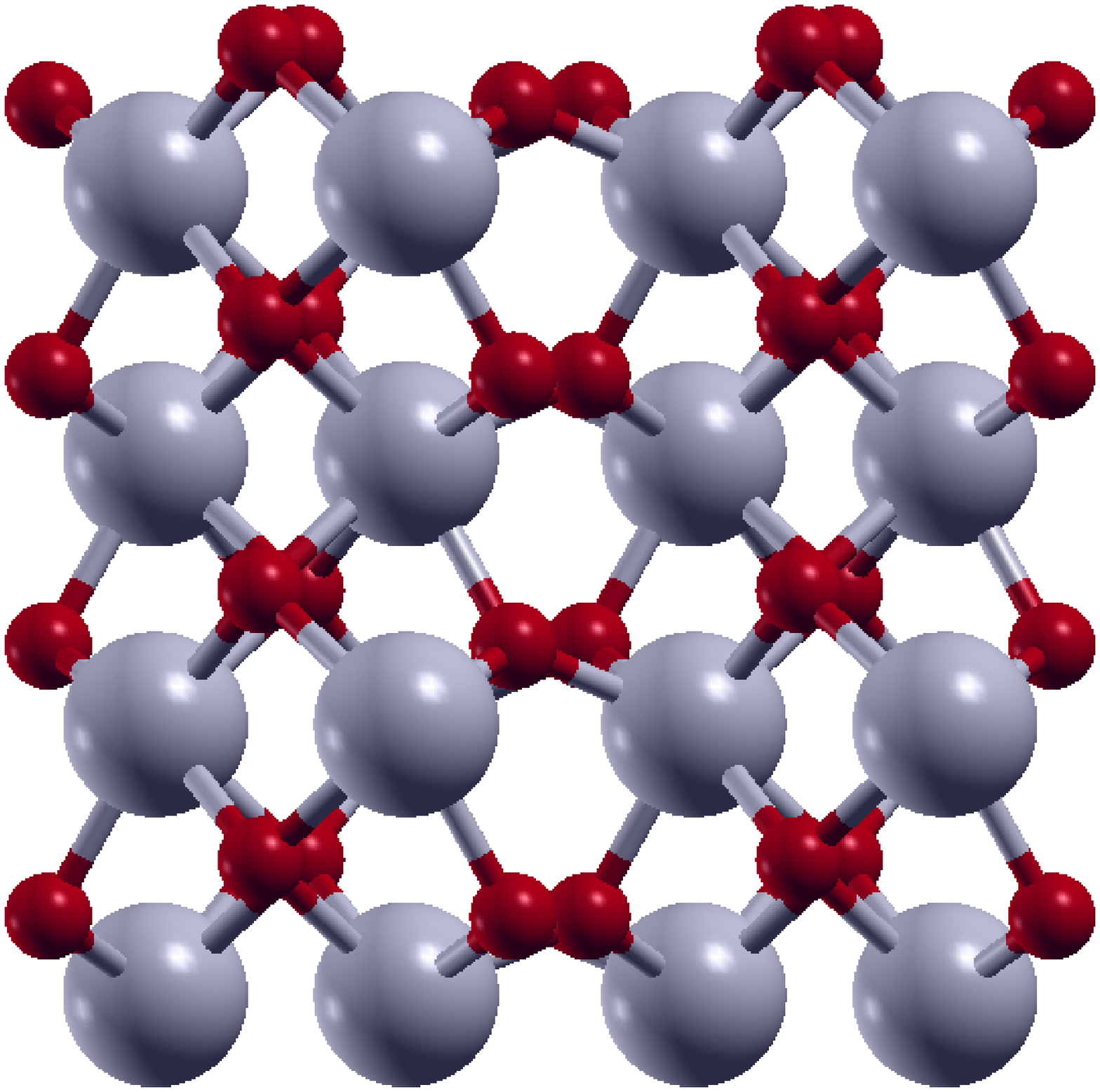}\hspace*{0.5cm}
\includegraphics[width=0.45\linewidth]{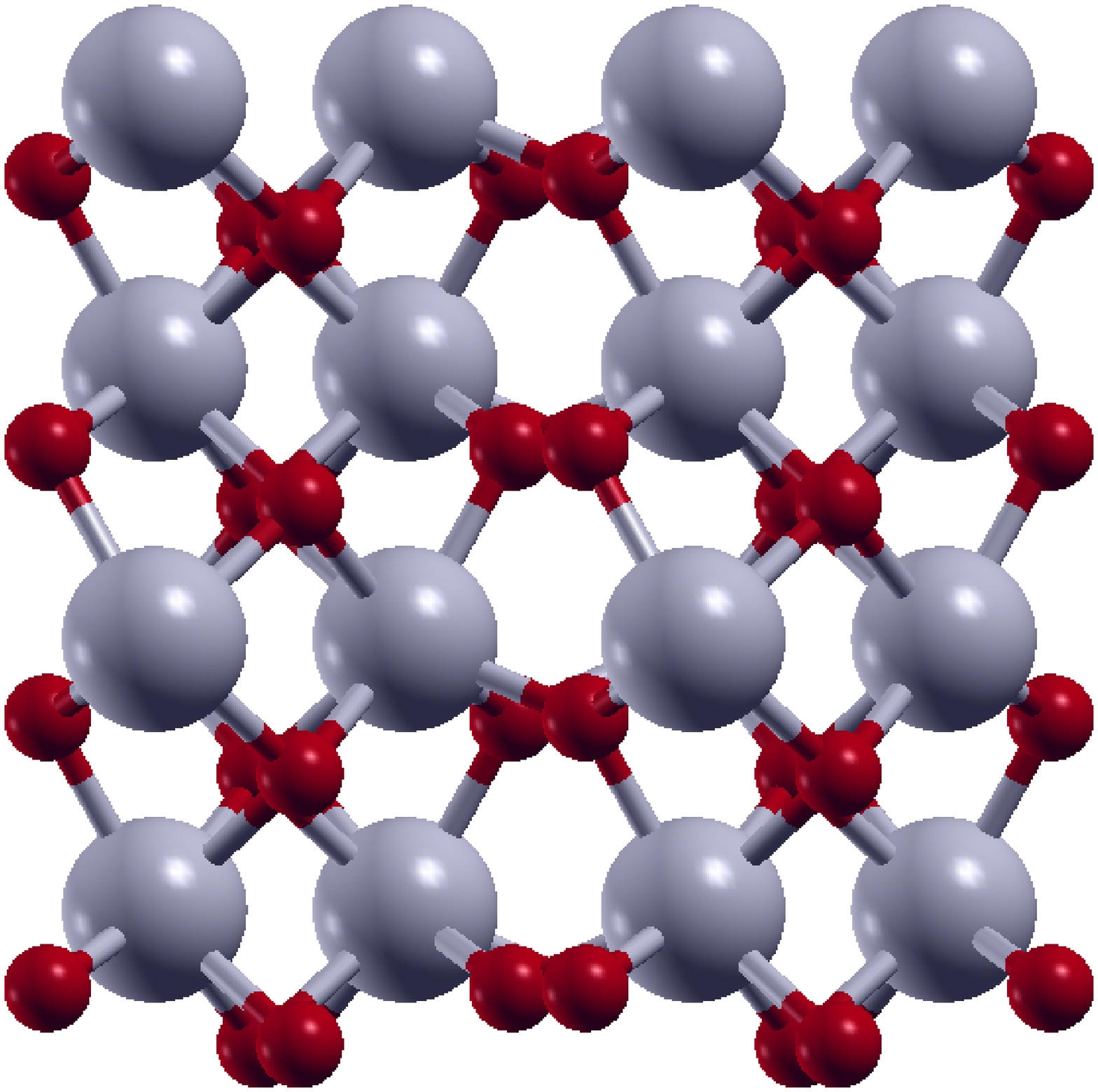}\\
\includegraphics[width=0.08\linewidth]{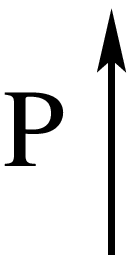}\hspace*{2cm}
\includegraphics[width=0.15\linewidth]{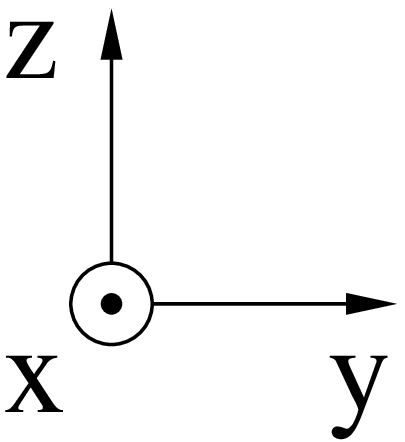}\hspace*{2cm}
\includegraphics[width=0.08\linewidth]{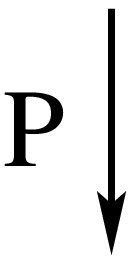}
\caption{Bulk ferroelectric HfO$_2$ structure in the orthorhombic phase with space group $Pca2_1$: view along $[\bar{1}00]$ direction depicting the $(y,z)$ plane of a $2\times2\times2$ supercell, for the two polarization configurations.}
\label{bulk}
\end{figure}

The unit cell contains 12 atoms, 4 Hf atoms and 8 O atoms. Starting from the Wyckoff positions of Hf1, O1 and O2 atoms \cite{doi:10.1063/1.4916707}, the atomic coordinates can be found using the four symmetry operations of the space group $Pca2_1$ \cite{PhysRevB.95.245141}. The four O1 atoms are 3-coordinated, while the four O2 atoms are 4-coordinated, having 3 or 4 neighboring Hf atoms, respectively. The polarization $P$ is oriented along the $[001]$ direction, with the opposite orientation for $[00\bar{1}]$. The polarization switching from $P$ to $-P$ can be understood from the displacement of the four O1 atoms along $z$-axis \cite{doi:10.1063/1.4867975}, as it is indicated in Fig.\ \ref{bulk}, where a view along $[\bar{1}00]$ is depicted. The atomic configurations of the bulk systems are shown in more detail in Fig.\ S1.        

\begin{figure}[t]
\centering
\includegraphics[width=0.45\linewidth]{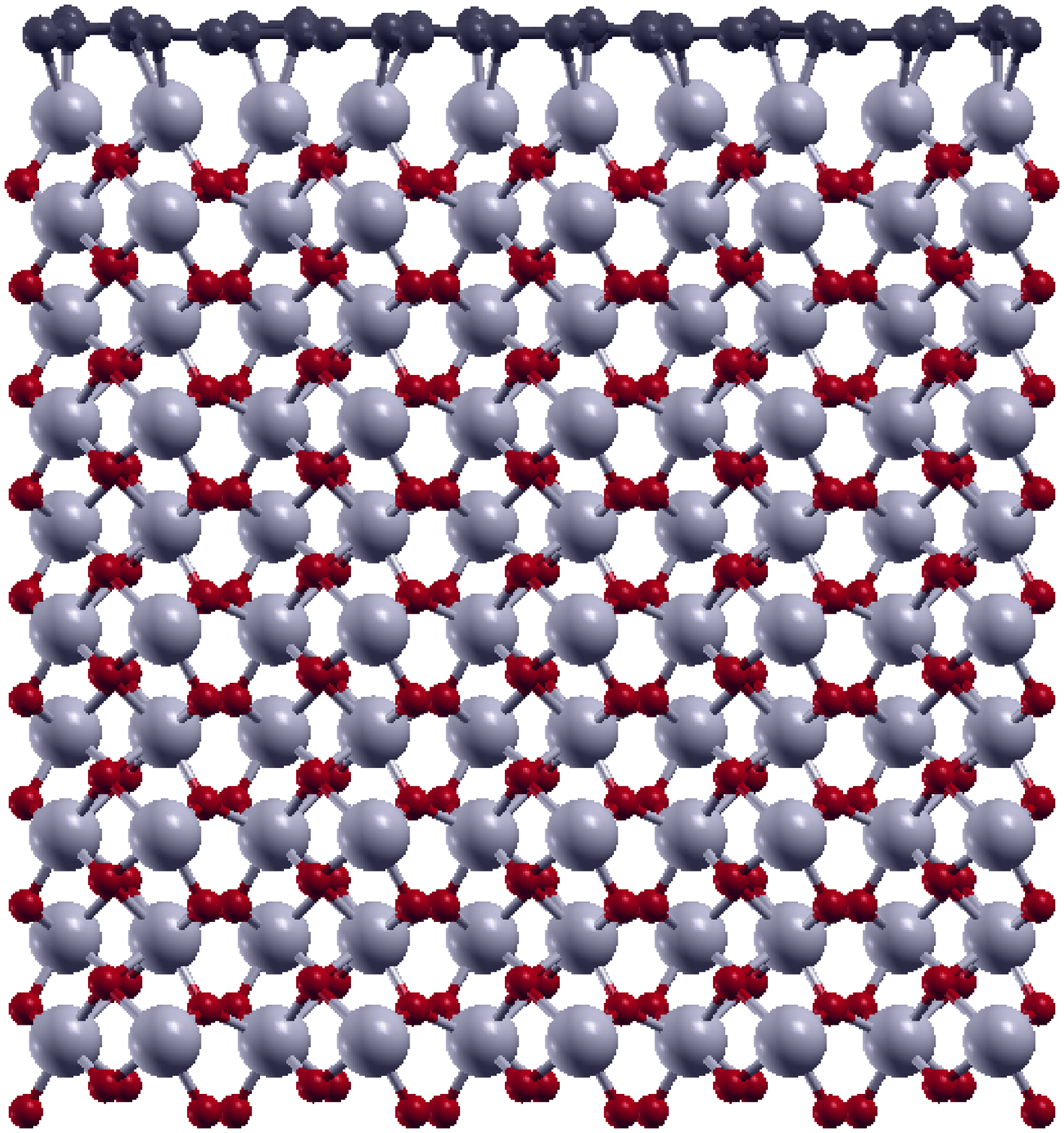}\hspace*{0.5cm}
\includegraphics[width=0.45\linewidth]{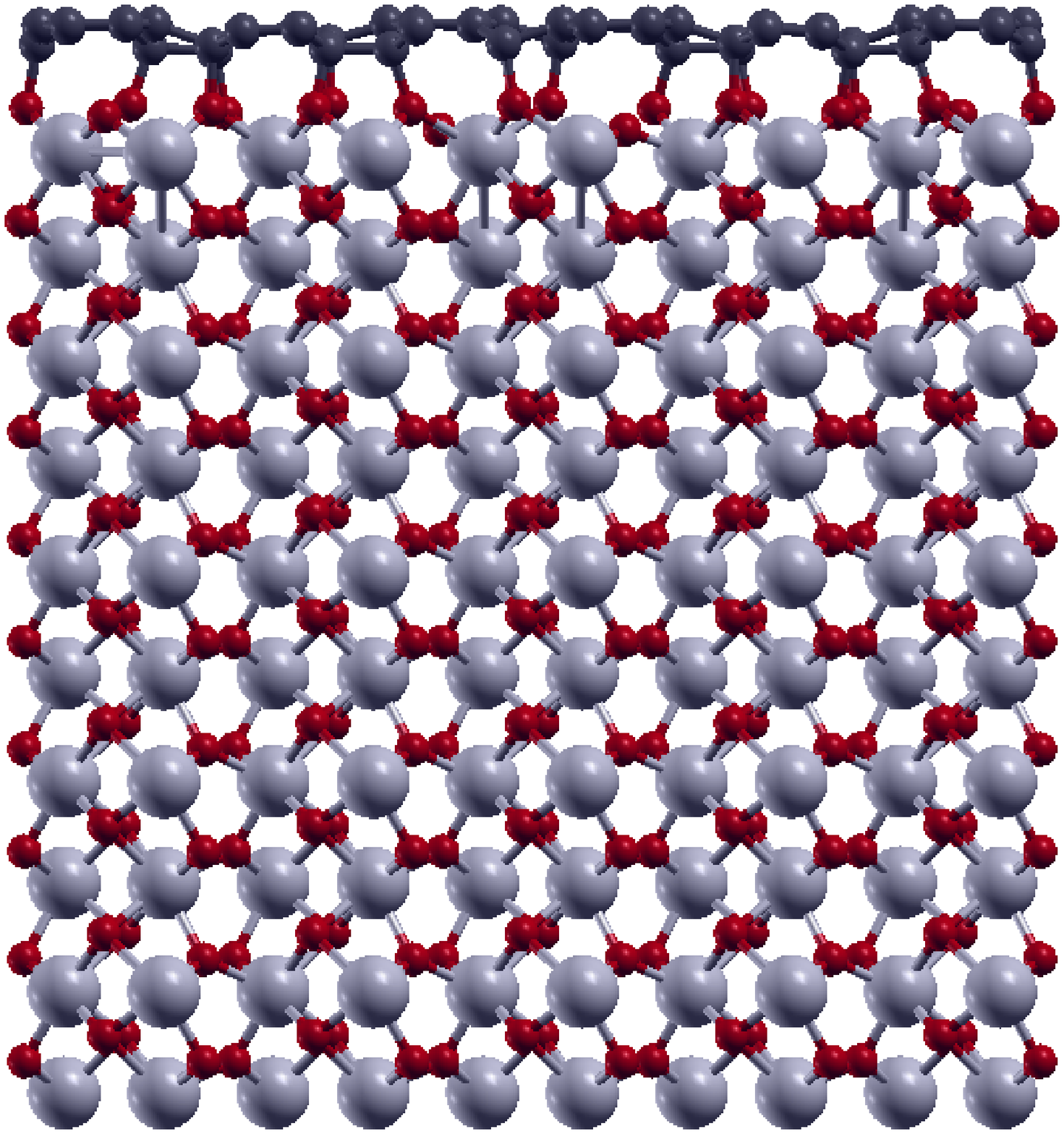}
(a) G@Hf-O \hspace*{3cm} (b) G@O-Hf
\caption{Graphene on top of (a) Hf- and (b) O-terminated HfO$_2$(001) surfaces, for the same polarization P pointing upwards. The oxygen layer present at the interface in G@O-Hf slab produces larger structural deformations in graphene compared to G@Hf-O structure.}
\label{structures}
\end{figure}

Using the optimized bulk parameters for both graphene and HfO$_2$, the interface systems are assembled. In order to minimize the stress introduced by the matching conditions, we consider a supercell of $1 \times 5 \times 5$ HfO$_2$ unit cells, on top of which the graphene layer is set. Both Hf- and O-terminations of the HfO$_2$ slab interacting with the graphene layer are considered, together with the two orientations of the polarization. We shall denote by G@Hf-O and G@O-Hf the structures with graphene on top of Hf and O layers, respectively, as shown in Fig. \ref{structures} and, from different angles, in Fig.\ S2. The interface systems totalize a number of 348 atoms, with 300 atoms in the HfO$_2$ slabs and 48 C atoms in graphene layer in the supercell. Two additional structures, with symmetrical HfO$_2$ slab terminations, denoted by G@Hf-Hf and G@O-O, have been also considered, as depicted in Fig.\ S3.

The DFT calculations are performed using the SIESTA package \cite{0953-8984-14-11-302}, which employs numerical atomic orbitals as basis set. For the bulk calculations the double-$\zeta$ polarized basis set was used and the real space grid was fixed by 300 Ry mesh cutoff. Norm-conserving Troullier and Martins pseudopotentials were employed with the following valence electron configurations: Hf -- 6s$^2$, 5d$^2$; O -- 2s$^2$, 2p$^4$; C -- 2s$^2$, 2p$^2$. For the exchange-correlation functional we use LDA in the parametrization proposed by Ceperley and Alder \cite{PhysRevLett.45.566}. As the energy gaps in HfO$_2$ are severely underestimated by LDA, we employ the LDA+U approach, setting on-site interactions for Hf 5d and O 2p orbitals. A k-point sampling scheme of $5\times5\times5$ was employed for the integrals in the 1BZ, while for generating the density of states (DOS) a finer grid of $20\times20\times20$ was used. The atomic coordinates are optimized until the residual forces are less than 0.04 eV/\AA. For the interface structures k-point sampling schemes of $3\times3\times1$ and $10\times10\times1$ were used for 1BZ integrals and DOS representation, respectively. Furthermore, the structural optimizations were first performed on thinner HfO$_2$ slabs and subsequently the number of layers was increased, up to 5 layers being considered. 

\section{Results and discussion}

\begin{figure}[t]
\centering
\includegraphics[width=0.95\linewidth]{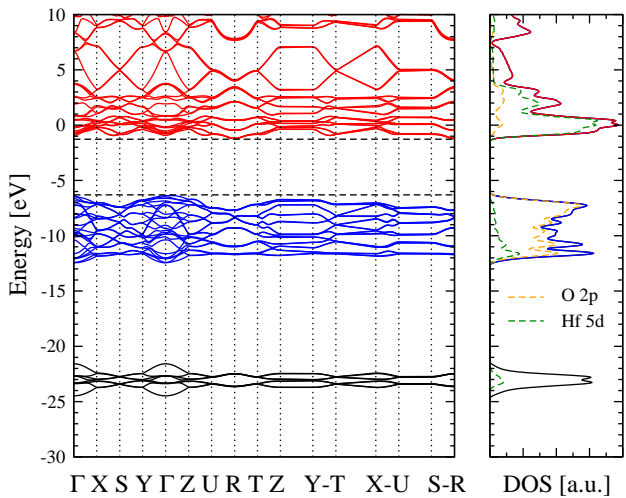}
\caption{Band structure and DOS of orthorhombic HfO$_2$ (space group $Pca2_1$). The obtained LDA+U gap is $E_{\rm g} = 5.03$ eV, marked by horizontal dashed lines. The valence band  maximum is located at $\Gamma$ point, while the conduction band minimum is found along the $R-T$ segment. Partial density of states corresponding to O-2p and Hf-5d orbitals is also represented. For plotting DOS, the broadening of the energy levels is 0.2 eV.}
\label{band-struct}
\end{figure}

Bulk structures corresponding to orthorhombic HfO$_2$ and graphene are first optimized. Starting with HfO$_2$ and taking the initial atomic coordinates from Ref.\ \cite{doi:10.1063/1.4916707} we obtain the optimized configuration, with lattice constants $a=5.17$ \AA, $b=5.05$ \AA\ and $c=5.07$\AA, which are close to the reported experimental values \cite{doi:10.1063/1.4919135}. Reproducing the correct band gap in several oxides by standard LDA is problematic, one notorious case being ZnO \cite{Bashyal_2018}. A semi-empirical approach based on DFT+U was successfully used to describe the electronic and optical properties of cubic HfO$_2$ \cite{LI2014397}. Here, we employ LDA+U and set the Coulomb on-site interactions $U^d = 8$ eV and $U^p = 1$ eV for Hf 5d and O 2p orbitals, respectively. The band structure is depicted in Fig.\ \ref{band-struct} along the selected k-paths. Our calculations confirm that HfO$_2$ is an indirect band gap semiconductor with LDA+U estimated gap energy $E_{\rm g} = 5.03$ eV. The top of the valence band located at $\Gamma$ point and the bottom of the conduction band is located long $R-T$ k-path \cite{PhysRevB.95.245141}. Although the LDA+U gap energy is still smaller than the experimental value of $\sim 5.7$ eV, it is considerably improved compared with the typical values obtained by plain DFT methods. The valence band is mostly formed by O 2p orbitals, while the conduction band largely consists of Hf 5d orbitals. Furthermore, using Berry phase approach we find a polarization of 68 $\mu$C/cm$^2$, which is similar to other previously reported values \cite{PhysRevB.90.064111,PhysRevB.95.245141}. Graphene structure is consistently reproduced at LDA level, with the inter-atomic distance matching the experimental value of 1.42\AA. 

Next, the interface structures are investigated starting from the bulk optimized parameters. As indicated in the previous section, we consider two types of systems, which correspond to Hf- and O-terminations interacting with graphene, labeled G@Hf-O and G@O-Hf, respectively. Unlike other ferroelectric materials, like e.g. Pb[Zr$_x$Ti$_{1-x}$]O$_3$, the HfO$_2$(001) oriented slab has polar surfaces originating from the charged top/bottom layers of Hf$^{+\delta}$ and O$^{-\delta}$. This contribution to the polarization is fixed by the growth conditions and it amounts to the switchable ferroelectric polarization. In this context, one should note that the presence of a base substrate for the HfO$_2$ thin film, typically Si or SiO$_2$, may further influence the charge distribution at the HfO$_2$ slab surfaces.

\begin{figure}[t]
\centering
\includegraphics[width=\linewidth]{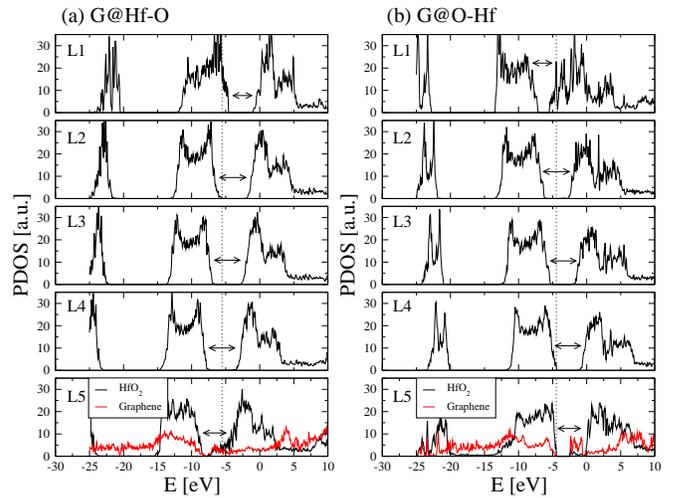}
\caption{Partial density of states for (a) G@Hf-O and (b) G@O-Hf systems. Contributions to the PDOS are indicated for each HfO$_2$ layer (black, L1-L5). In the bottom sub-plots the graphene PDOS (red) is represented, revealing energy gaps, which correlate well with the reduced bulk-like gap of the HfO$_2$ L5 interfacial layer. The Fermi level is marked by vertical dotted lines. }
\label{slab-pdos}
\end{figure}

Figure \ref{slab-pdos} shows the partial density of states (PDOS) for G@Hf-O and G@O-Hf systems for the same orientation of the polarization. The PDOS contributions are determined for each of the five layers of HfO$_2$ (L1-L5), each containing 60 atoms, and for the graphene layer, separately. One may observe that the PDOS contribution of each layer resembles well the bulk DOS, particularly the inner layers (L2,L3,L4). As expected, the two outer layers, the bottom layer (L1) and top layer (L5), neighboring vacuum and graphene, respectively, present a somewhat disturbed DOS, as some electronic states are located within the gaps. For both types of systems, the energy gaps of HfO$_2$ layers are shifted on the energy axis, which is indicative of internal polarization of the slab. The energy shifts are in opposite directions: for G@Hf-O the top of the valence band is shifted downwards from L1 to L5 with $\sim 2.8$ eV, while for G@O-Hf we observe a shift of $\sim 2.5$ eV towards higher energies. In both systems, this corresponds to the charging of the outer atomic Hf layers by $+\delta$ and of the O layers by $-\delta$, with $\delta>0$. The effect of the polar surfaces was further investigated by considering, for comparison, two additional systems, G@Hf-Hf and G@O-O, where the top and the bottom atomic layers are the same, either Hf or O, respectively. The structures are depicted in Fig.\ S3 and the layer-by-layer PDOS is indicated in Fig.\ S4. In this case, there is no significant shift in the band gaps corresponding to layers L1-L5. 

\begin{figure}[t]
\centering
\includegraphics[width=0.9\linewidth]{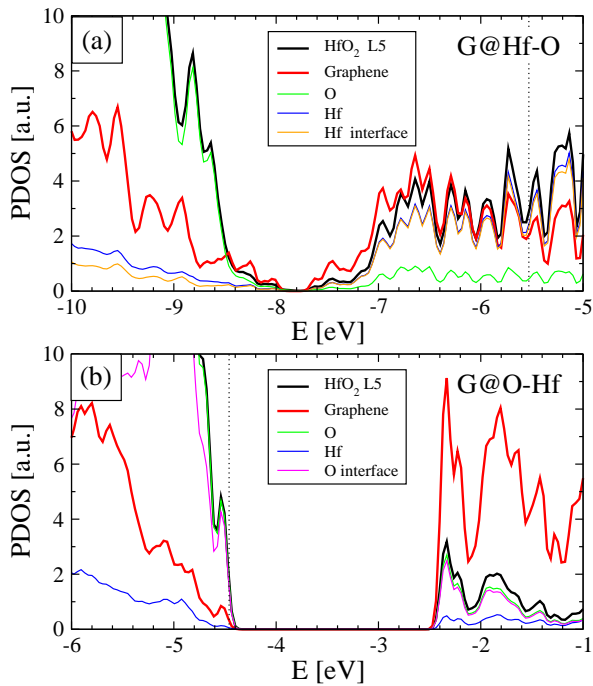}
\caption{Detailed PDOS contributions for (a) G@Hf-O and (b) G@O-Hf systems, corresponding to the interface HfO$_2$ L5 layer and graphene. Individual contributions from Hf and O atoms in HfO$_2$ L5 layer are represented and the Fermi level is represented by vertical dotted lines. The contributions from the interfacial Hf atoms (a) and from the O atoms (b) are shown to be dominant at the gap edges.}
\label{pdos-details}
\end{figure}

The band gaps observed in the PDOS corresponding to HfO$_2$ L5 layers are significantly smaller than in the bulk, $\sim 0.25$ eV for G@Hf-O and $\sim 1.8$ eV for G@O-Hf. Importantly, however, is the appearance of energy gaps in graphene PDOS precisely at the same location as in the HfO$_2$ L5 layers. In fact, the PDOS associated with the orbitals of carbon atoms seems to follow closely the PDOS around the gap edges of the HfO$_2$ interfacial layers. The Fermi levels, shown by vertical doted lines, are crossing the conduction band edge and valence band edge corresponding to the L5 layers, for G@Hf-O and G@O-Hf structures, respectively. Instead, for the inner layers (L2,L3,L4), which correspond to the bulk of HfO$_2$ slabs, the Fermi energy lies close to the middle of the gap, as in a typical intrinsic semiconductor. In systems with symmetric terminations, G@Hf-Hf and G@O-O, we observe similar gaps, as it is shown in Fig.\ S4. It is worth noting that for the G@O-O system the Fermi level is shifted into the gap. Therefore, the transport properties in a graphene based FET may sensibly depend on the termination types of HfO$_2$ slabs and also on the base substrate. Thus, in experimentally relevant conditions, the position of the Fermi level may be influenced by doping, the nature of the substrate and/or by the application of a gate voltage. Looking at the PDOS stemming only from the interfacial Hf atoms, we notice that their contribution is significant for the lower part of the conduction band. In fact, the Hf 5d states of the interfacial atoms effectively reduce the bulk-like gap at the interface.

\begin{figure}[t]
\centering
\includegraphics[width=\linewidth]{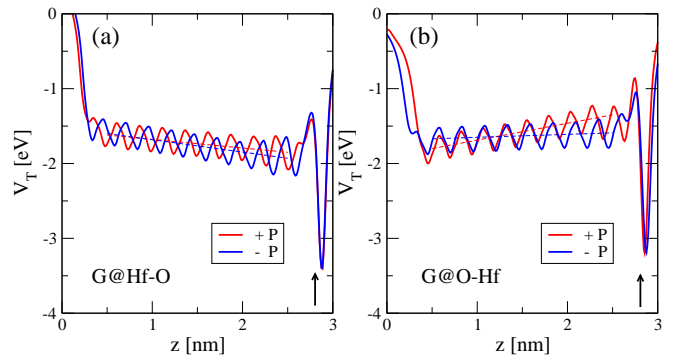}
\caption{Total potential energy $V_{\rm T}(z)$, averaged in the $(x,y)$ plane for (a) G@Hf-O and (b) G@O-Hf, showing the effect of the polar Hf$^{+\delta}$ and O$^{-\delta}$ surfaces. For each system, $V_{\rm T}$ is represented for the two ferroelectric polarizations, with the dashed lines showing the changes in the polarization fields. The arrows mark the approximate positions of the graphene layers.}
\label{VT}
\end{figure}

Inducing energy gaps in graphene is crucial for FET applications and this is typically achieved by symmetry breaking between the A and B type sublattices, e.g. using BN or SiC substrates \cite{PhysRevB.76.073103,Zhou2007}. However, here there is no direct matching of graphene and HfO$_2$(001) surface, along the lines of the hexagonal symmetry. Instead we observe the locally deformed graphene structure, which is enhanced for the O-terminated surface of G@O-Hf system. In order to assess the formation of the energy gaps we examine in more detail the PDOS contributions in Fig.\ \ref{pdos-details}. Analyzing the G@Hf-O system in Fig.\ \ref{pdos-details}(a), the valence band edge mostly consists of O orbitals, while the conduction band is composed by Hf orbitals, with significant contributions from $p$ and $d$ shells, respectively, as in the bulk case. However, in the L5 layer of the G@Hf-O system, the conduction band edge is shifted, due to the intercalation of in-gap states, leaving the small gap of $\sim 0.25$ eV, which is also visible in the graphene PDOS. The interfacial Hf atoms and their interaction with graphene are mainly responsible for in-gap states present in the bulk-like gap. The resemblance between the PDOS spectra of both graphene and Hf atoms is indicative for the orbital hybridization, particularly between $\pi$ orbitals of graphene and $d$ orbitals of Hf, as already reported in Ref. \cite{PhysRevB.83.153413} for cubic hafnia. Considering now the G@O-Hf system, a significantly larger gap of $\sim 1.8$ eV is present in the interfacial HfO$_2$ L5 layer. Here, in contrast to G@Hf-O system, the bottom of the conduction band is formed by O orbitals as in-gap states, and mostly by the O atoms present at the interface. Again, at the bottom of the conduction band the graphene PDOS is matching well contributions from interfacial O atoms and Hf atoms in L5.

In addition we explore ferroelectric polarization switching effects. Figure\ \ref{VT} shows the total potential energy $V_{\rm T}(z)$ as a sum of the local pseudopotential, Hartree and exchange-correlation potentials, averaged in the $(x,y)$ plane. For the G@Hf-O system, $V_{\rm T}$ drops towards the graphene layer, while it increases in the case of G@Hf-O in accordance with the previous PDOS analysis. Taking into account the two orientations of the ferroelectric polarization the potential energy maps are slightly tilted around the direction imposed by the polar surfaces. The estimated polarization field is in the range of $0.2-0.9$ MV/cm. The HfO$_2$ layers are identifiable from the maxima in the potential energy, while the graphene layer is positioned at $z \simeq 2.8$ nm. This analysis shows that the polar nature of the HfO$_2$ slab surfaces set by the termination types can determine charge accumulation or depletion at interface with graphene, creating interfacial dipoles.

Moreover, controlling the growth conditions for the ultra-thin HfO$_2$ is essential to obtain the desired field effect control in graphene. To this end, our results concerning G@O-Hf and G@O-O systems suggest that oxygen-rich conditions are more suitable to achieve this goal by introducing significant energy gaps in graphene. Furthermore, the Fermi energy may be properly adjusted, in particular by the choice of the base substrate.

Although in very thin HfO$_2$ slabs the polarization switching may not have a large impact on the electronic band structure it may influence influence the transport properties by at least two reasons. First, the switchable polarization field determines the electron density at the graphene layer, particularly in the context of induced energy gaps. Secondly, as the interactions between graphene and the HfO$_2$ substrate were identified as the cause for the diminished carrier mobility, the observed local deformations can be influenced by the polarization state and may actually tune the mobility of graphene. 


\section{Conclusions}

We investigated the electronic properties of graphene on top of ferroelectric HfO$_2$ with space group $Pca2_1$, using DFT calculations. Hf- and O-terminated slabs interacting with graphene were considered. Structural relaxations indicate local deformations of the graphene monolayer, which are enhanced for O-terminated slabs compared to Hf-terminated ones. The PDOS analysis reveals the appearance of energy gaps in graphene, which is a crucial aspect in the development of graphene based FETs on ferroelectric hafnia. In addition, ferroelectric polarization switching amounts to the local charging effects introduced by polar HfO$_2$(001) surfaces. We showed that the graphene induced band gaps are robust, independent on the type of the HfO$_2$ slab terminations, symmetric or asymmetric. Our analysis suggests that oxygen-rich growth conditions may be beneficial for inducing significant gaps, while the Fermi energy can be properly adjusted by the base substrate. Moreover, the local deformations induced by the oxygen interfacial layer in connection with polarization switching may determine electron mobility tuning in the graphene layer, which can directly influence the transport properties in graphene-on-hafnia based FETs.   \\

{\bf Author contributions}\\

G.A.N. developed the computational experiment and performed the computations. All authors discussed the results and commented on the manuscript.\\ 

{\bf Acknowledgements}\\

The authors acknowledge the financial support of project GRAPHENEFERRO grant of Ministry of Research and Innovation, CNCS-UEFISCDI, number PN-III-P4-ID-PCCF-2016-0033.



\bibliography{manuscript}

\onecolumngrid

\newpage 

\appendix*
\section{Supplementary Material}

\renewcommand{\thefigure}{S\arabic{figure}}
\setcounter{figure}{0}

\begin{figure*}[h]
\centering
\includegraphics[height=1.cm]{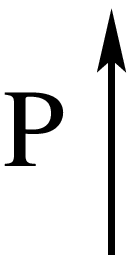} \hspace*{0.5cm}
\includegraphics[height=3.5cm]{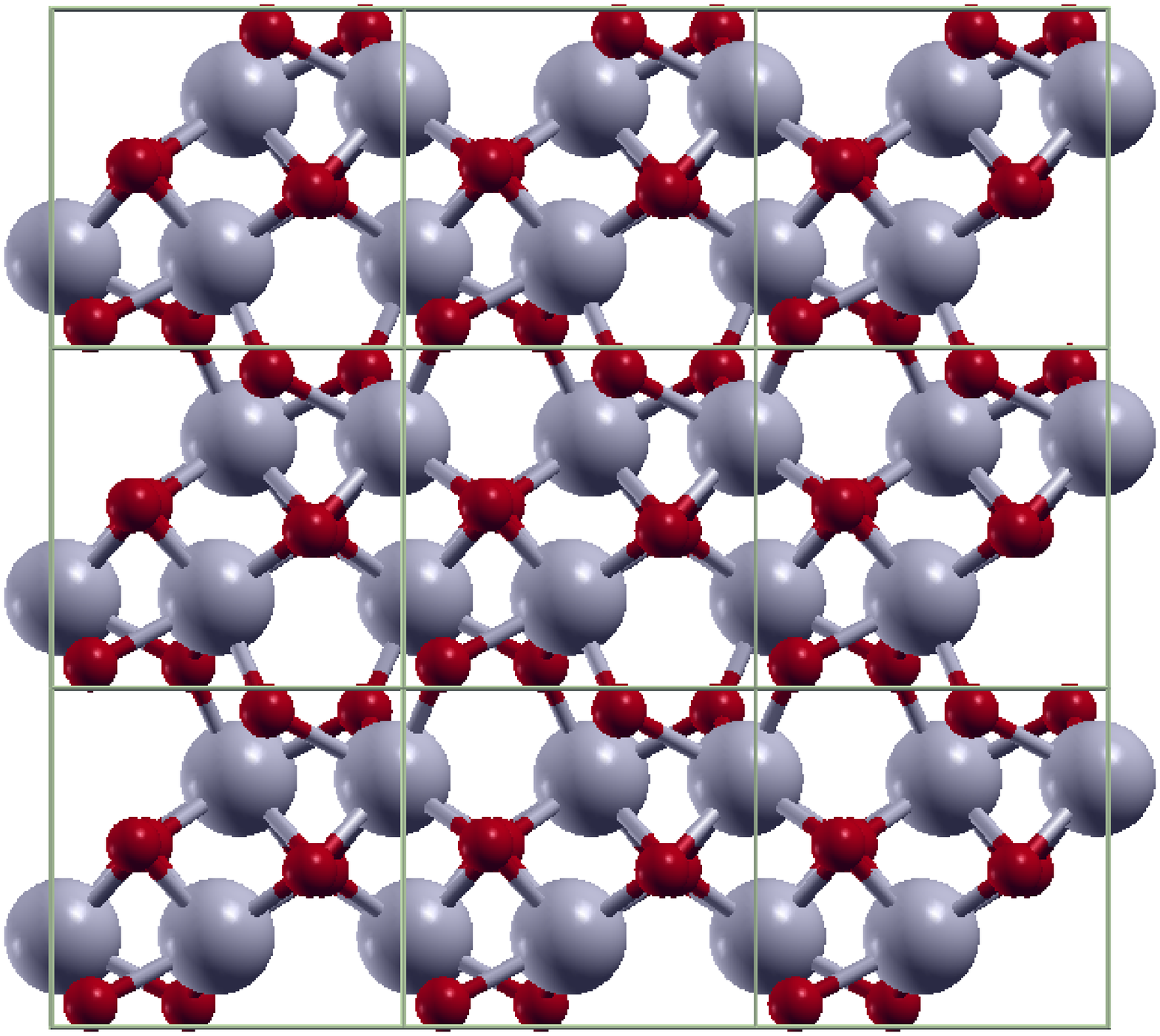} \hspace*{0.5cm}
\includegraphics[height=3.5cm]{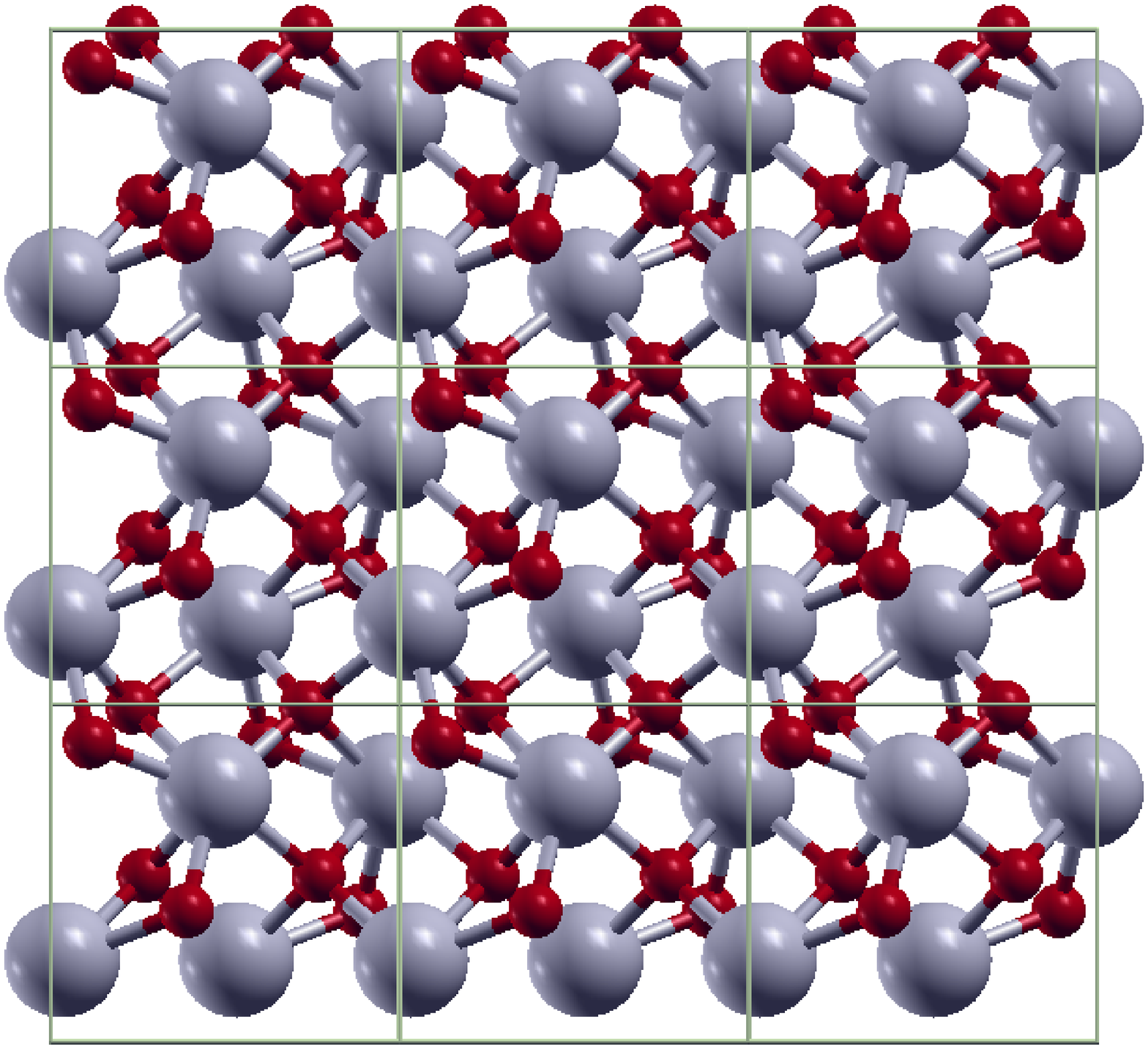} \hspace*{0.5cm}
\includegraphics[height=3.5cm]{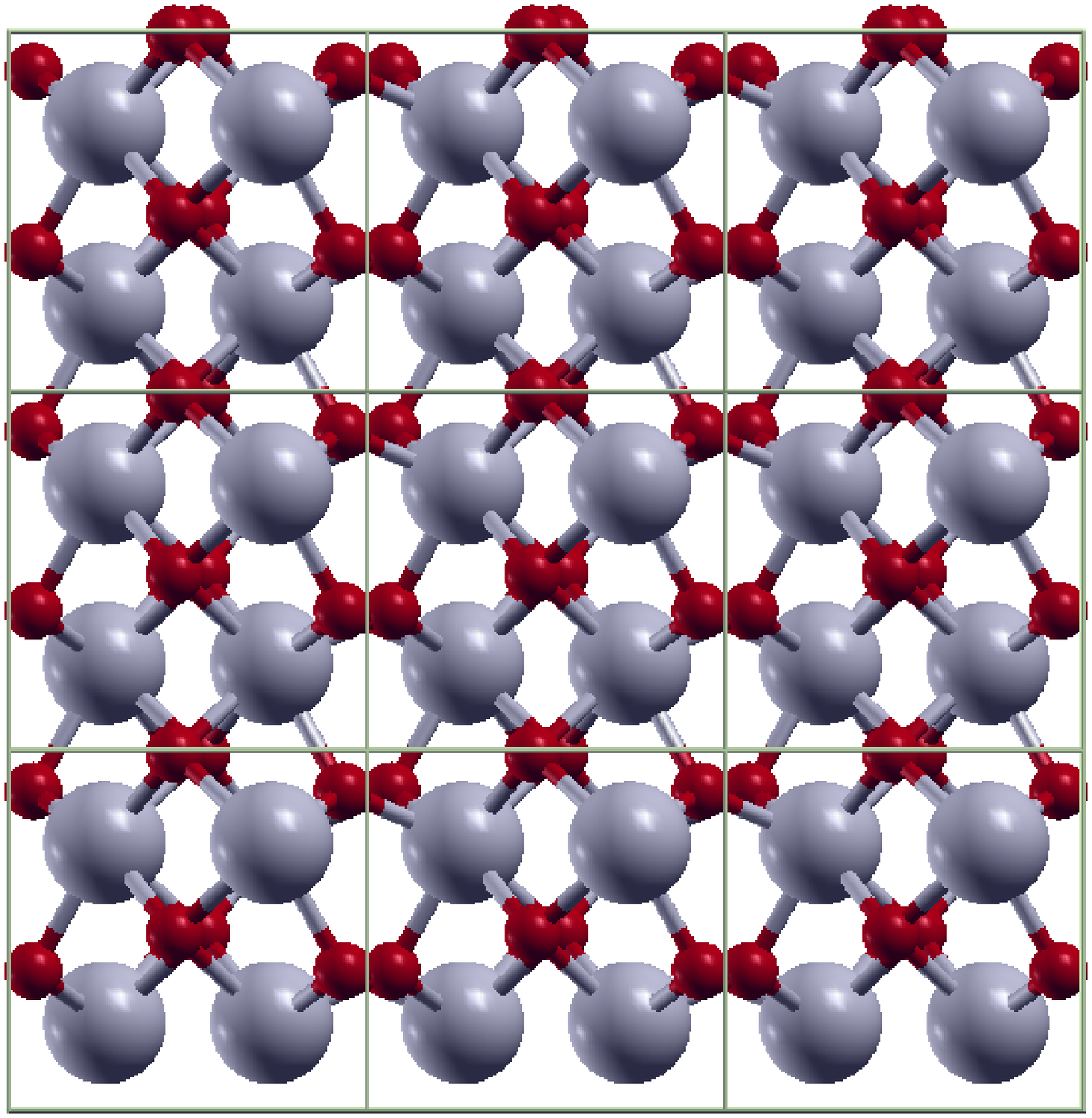} \\
\includegraphics[height=1.5cm]{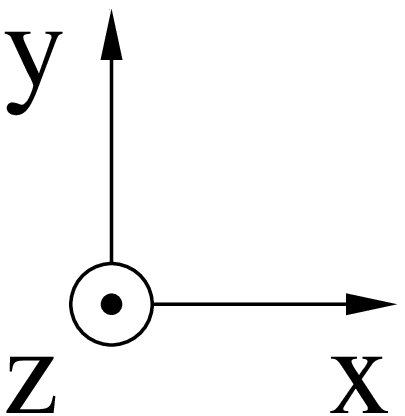} \hspace*{3.0cm}
\includegraphics[height=1.5cm]{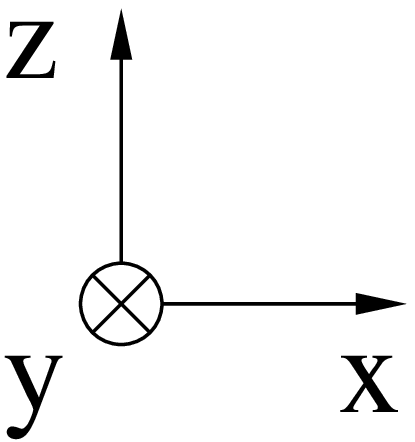} \hspace*{3.0cm}
\includegraphics[height=1.5cm]{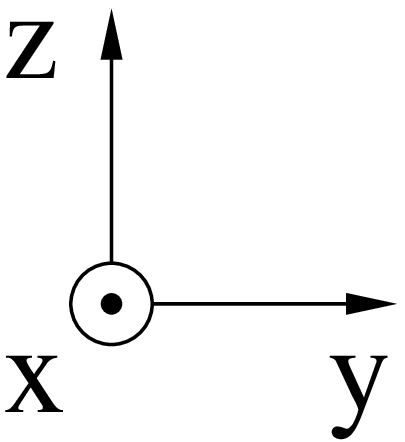} \\
\includegraphics[height=1.cm]{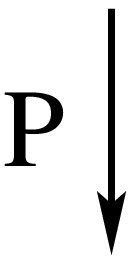} \hspace*{0.5cm}
\includegraphics[height=3.5cm]{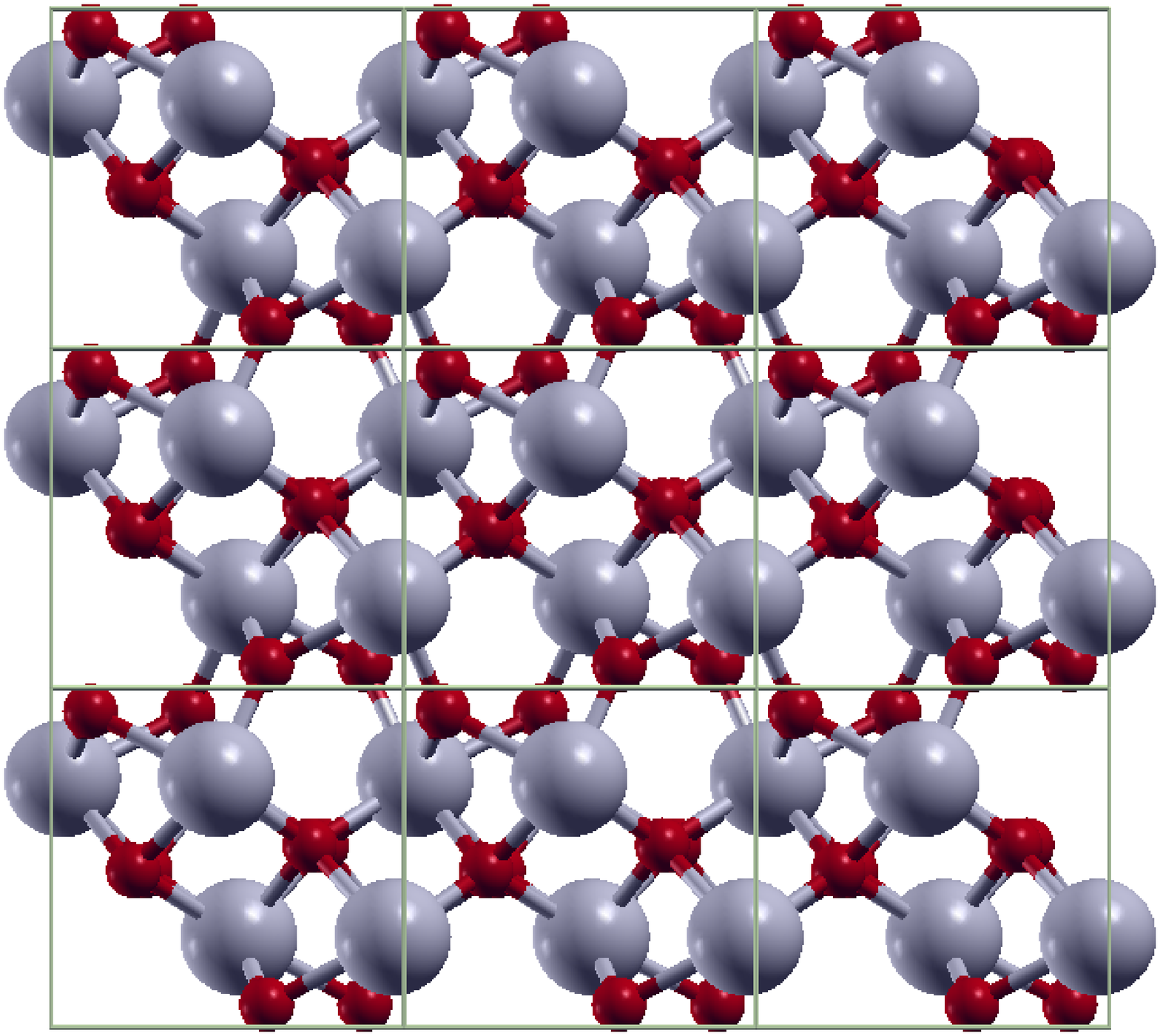} \hspace*{0.5cm}
\includegraphics[height=3.5cm]{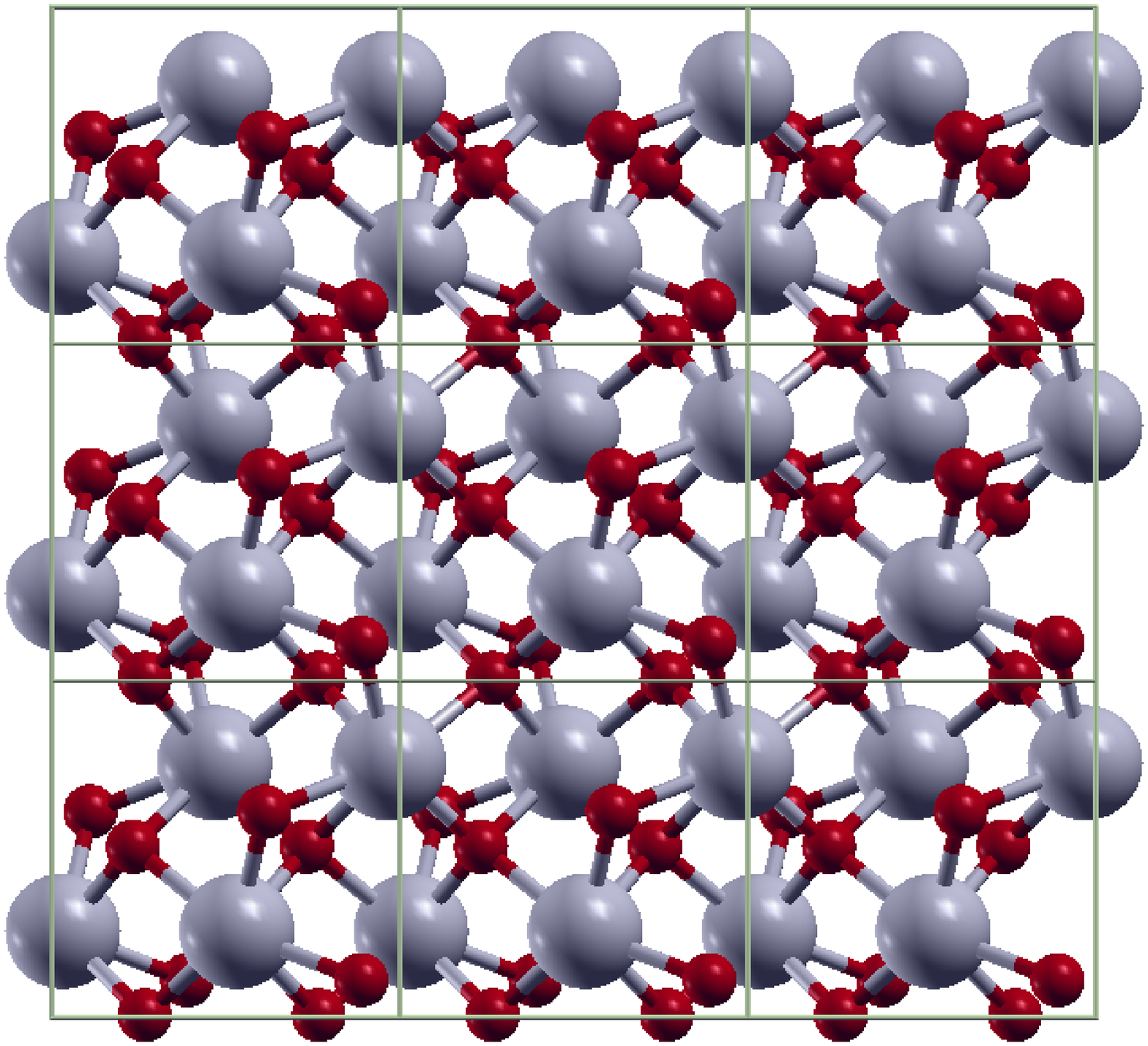} \hspace*{0.5cm}
\includegraphics[height=3.5cm]{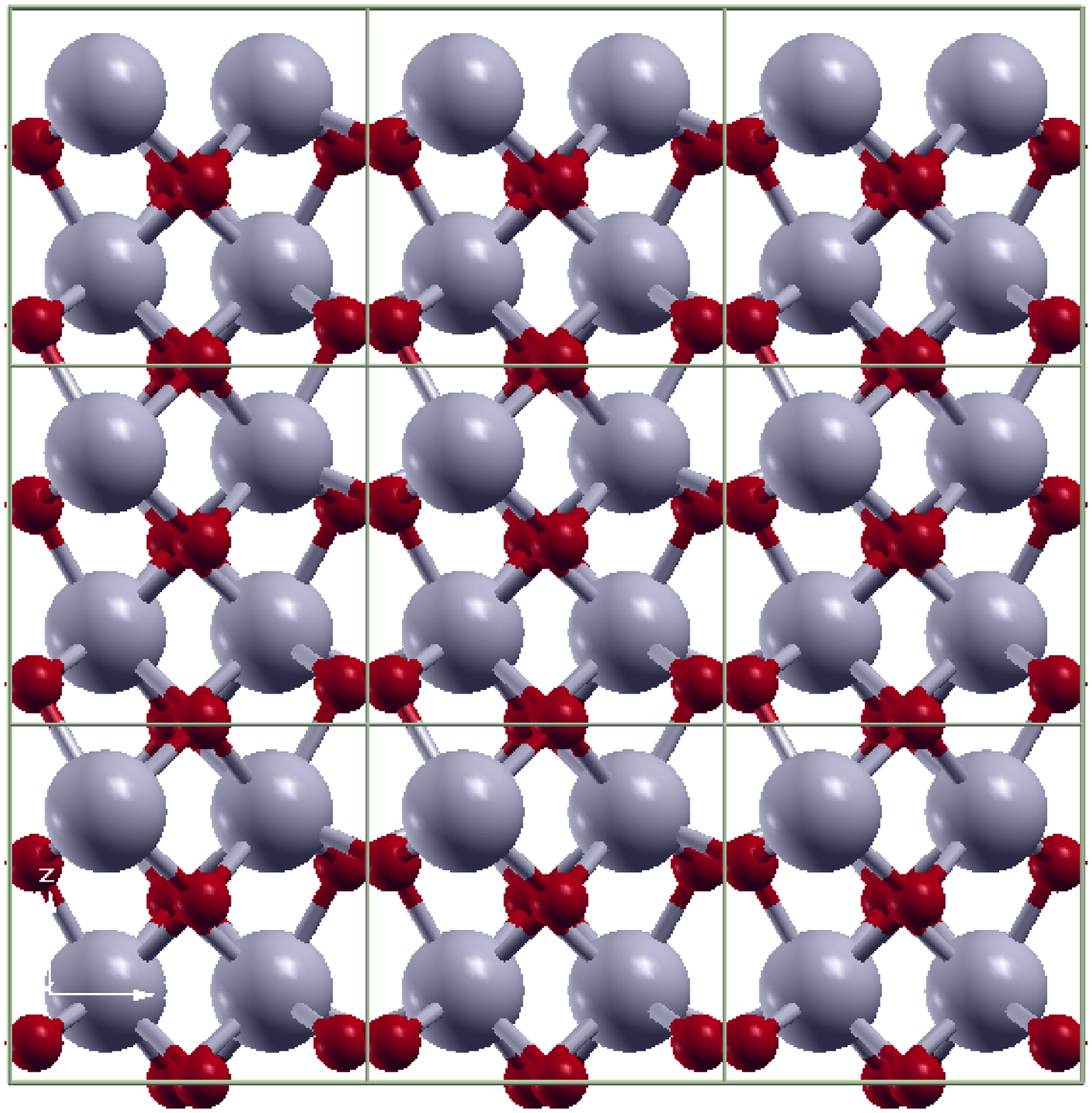} 
\caption{Atomic configurations of bulk HfO$_2$ with space group Pca2$_1$, exposing $(x,y)$, $(x,z)$ and $(y,z)$ faces, for $P$ and $-P$ polarizations. Color codes: Hf -- gray, O-- red.}
\label{bulk-struct}
\end{figure*}

\begin{figure*}[h]
\centering
\includegraphics[height=5.5cm]{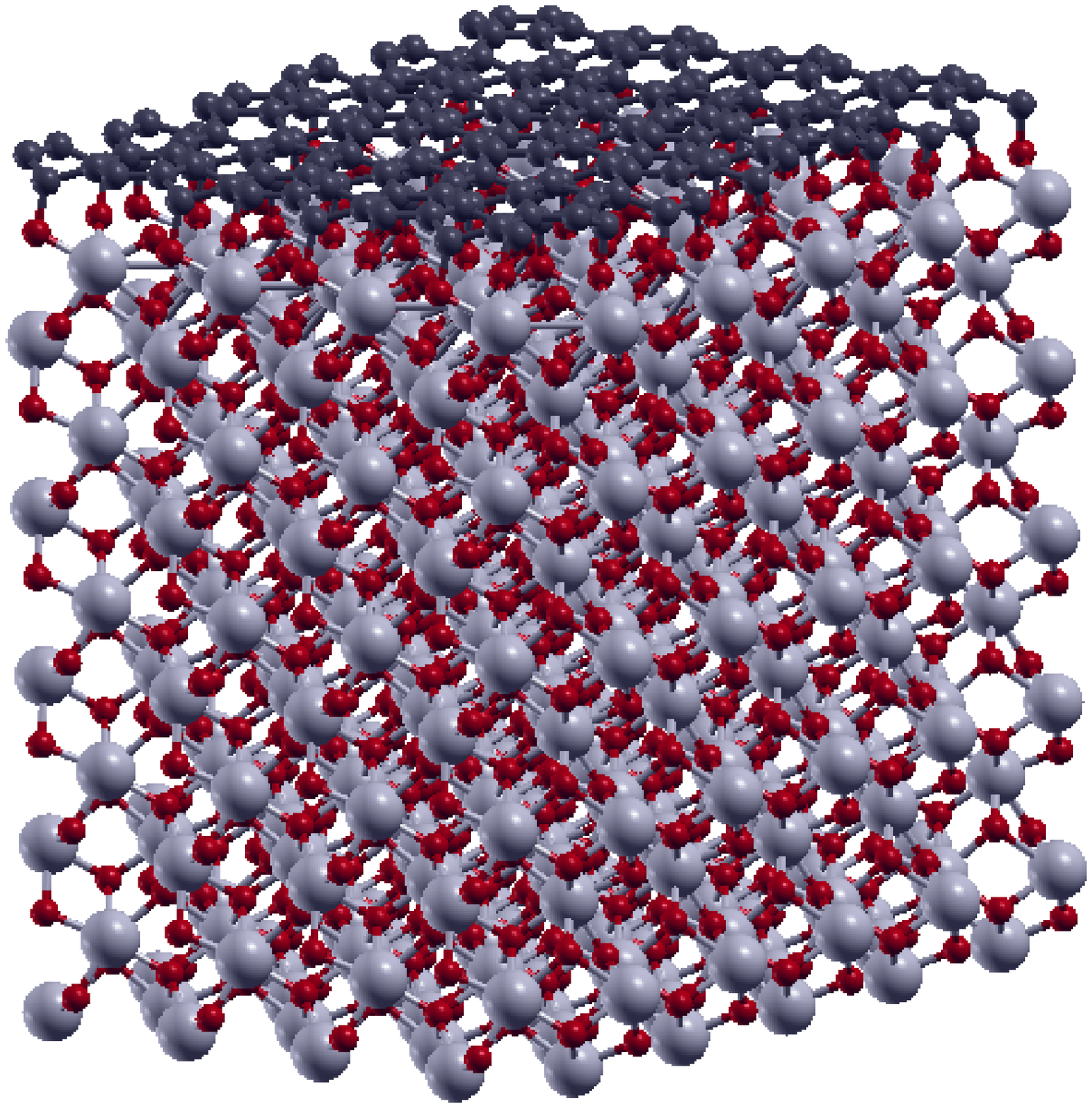} \hspace*{0.5cm}
\includegraphics[height=5.5cm]{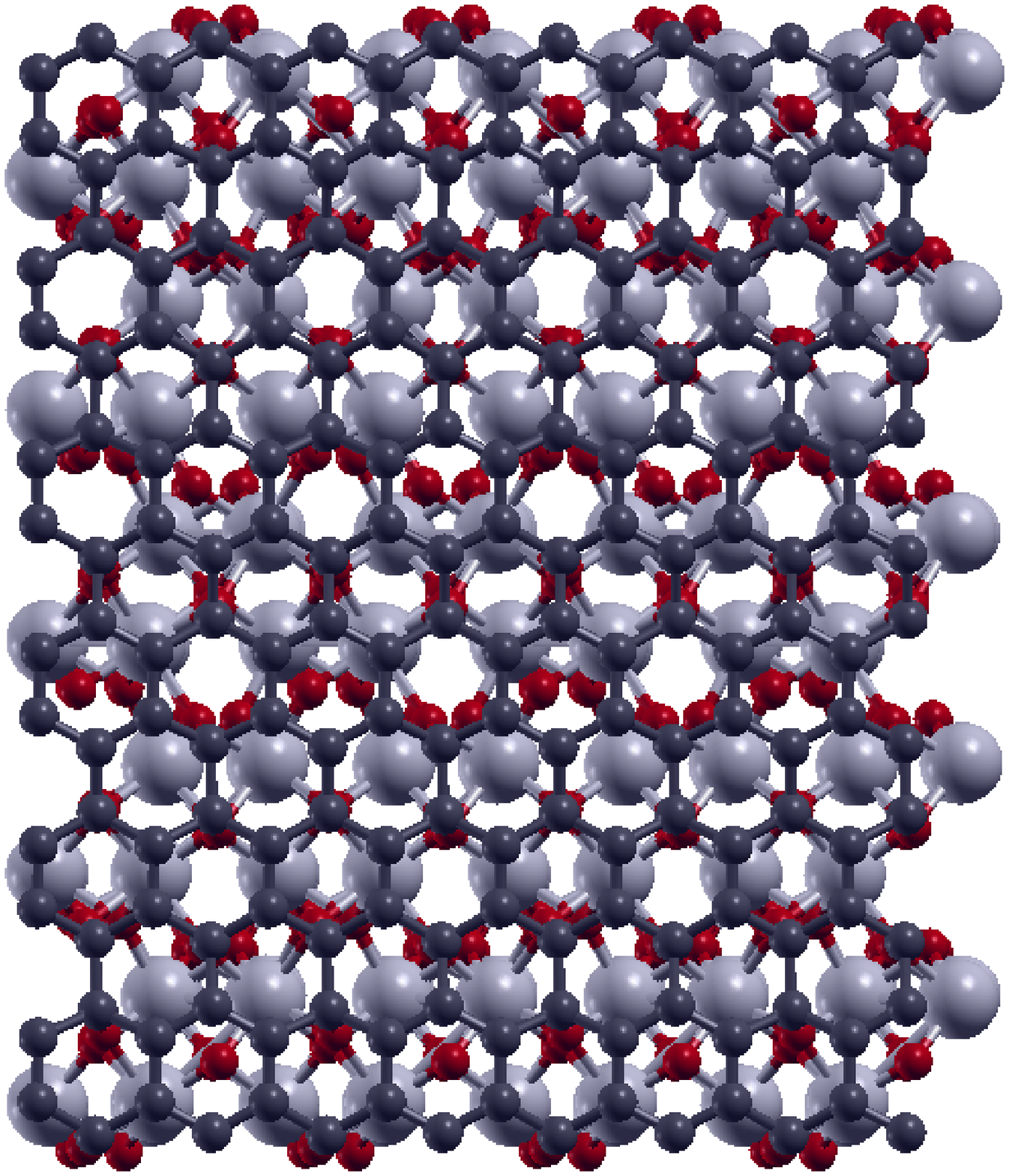} \\
(a) \hspace*{5cm} (b)
\caption{Graphene on top of the HfO$_2$ slab, depicting G@O-Hf system: (a) 3D view and (b) top view. In this case, graphene interacts with the O-terminated HfO$_2$ slab.}
\label{O-Hf_3D_toview}
\end{figure*}

\newpage

\begin{figure}[t]
\centering
\includegraphics[height=5.cm]{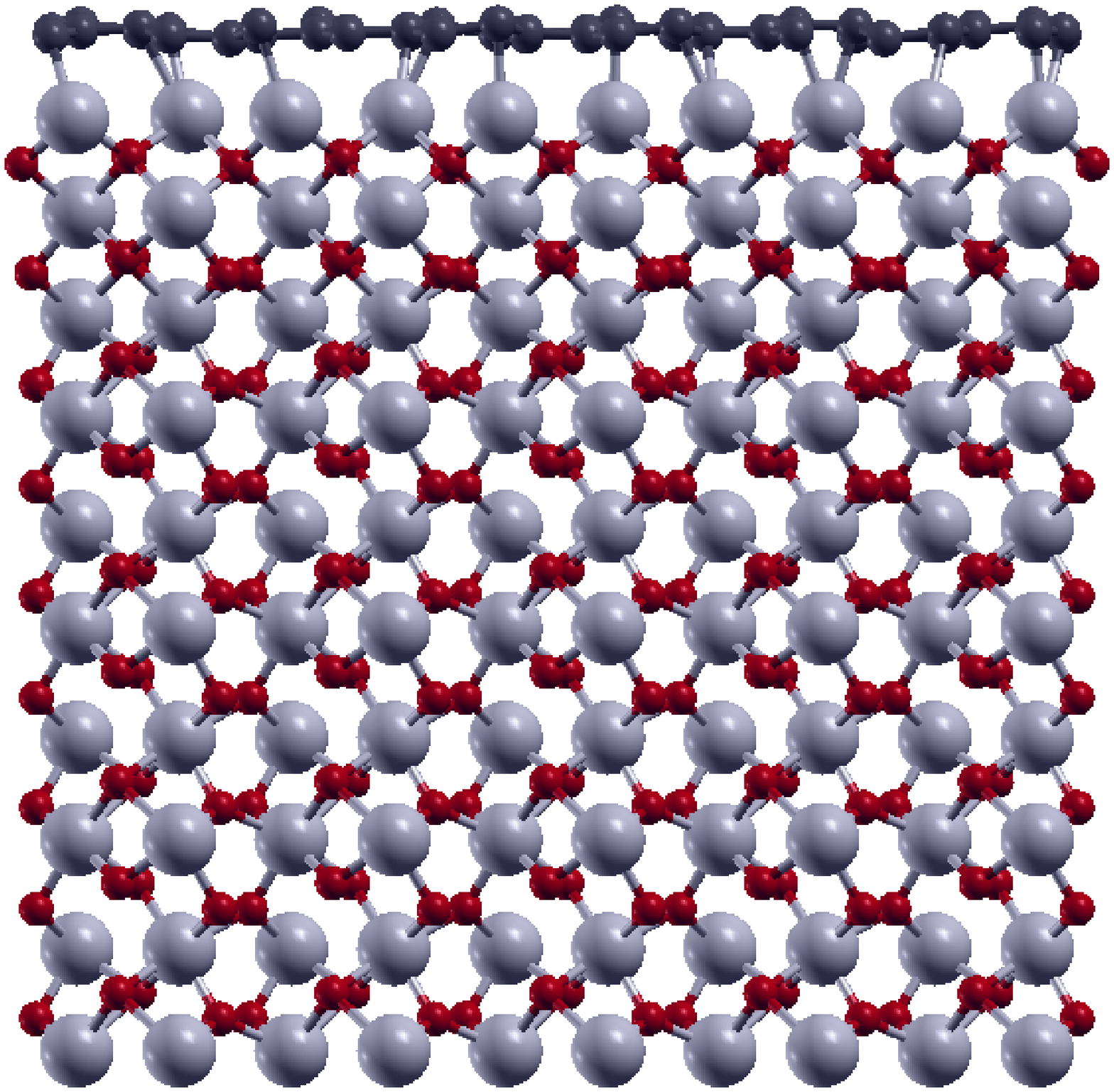} \hspace*{0.5cm}
\includegraphics[height=5.cm]{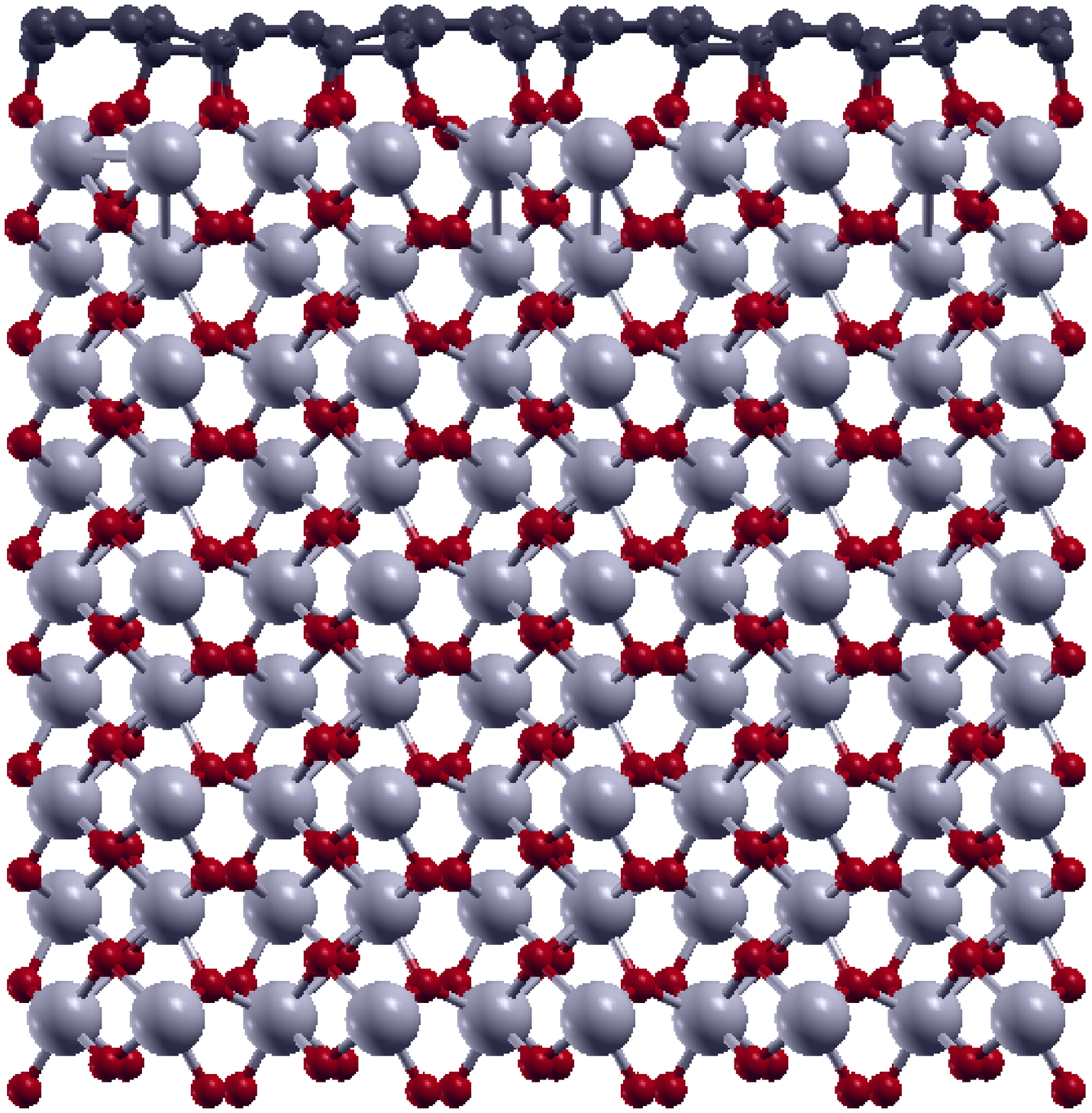} \\
(a) \hspace*{5cm} (b)
\caption{Graphene on top of HfO$_2$ slabs with symmetric terminations: (a) G@Hf-Hf and (b) G@O-O.}
\label{struct_Hf-Hf_O-O}
\end{figure}

\begin{figure}[t]
\centering
\includegraphics[height=7.5cm]{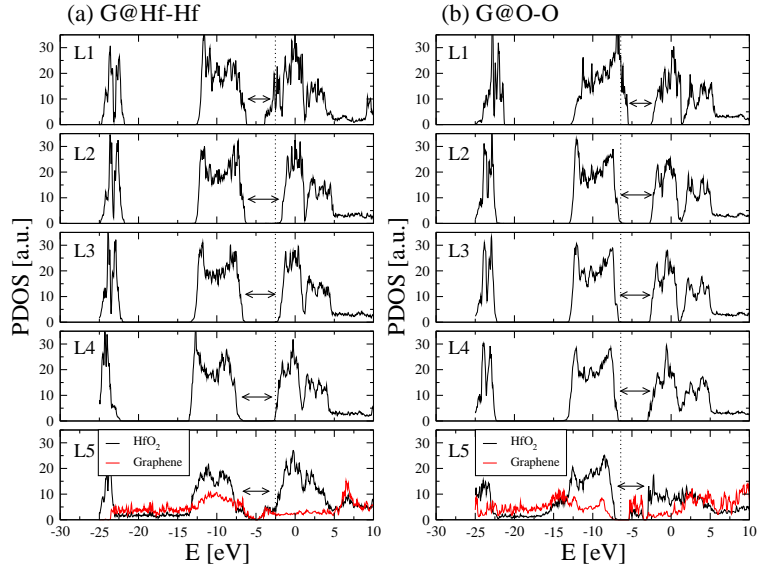} 
\caption{Partial density of states (PDOS) for (a) G@Hf-Hf and (b) G@O-O. Compared to the asymmetrically terminated slabs, G@Hf-O and G@O-Hf, the large shift induced by the polar surfaces is not anymore present. However, there is a similar behavior in PDOS for the interfacial HfO$_2$ layer and graphene, concerning the energy gaps. One notable difference is found in the slightly shifted Fermi level compared to the G@O-Hf system, which now enters the gap, suggesting that the bottom terminations of the HfO$_2$ slab or the base substrate can influence the transport properties of graphene.}
\label{PDOS_Hf-Hf_O-O}
\end{figure}

\newpage

\begin{figure*}[h]
\centering
\includegraphics[height=10.cm]{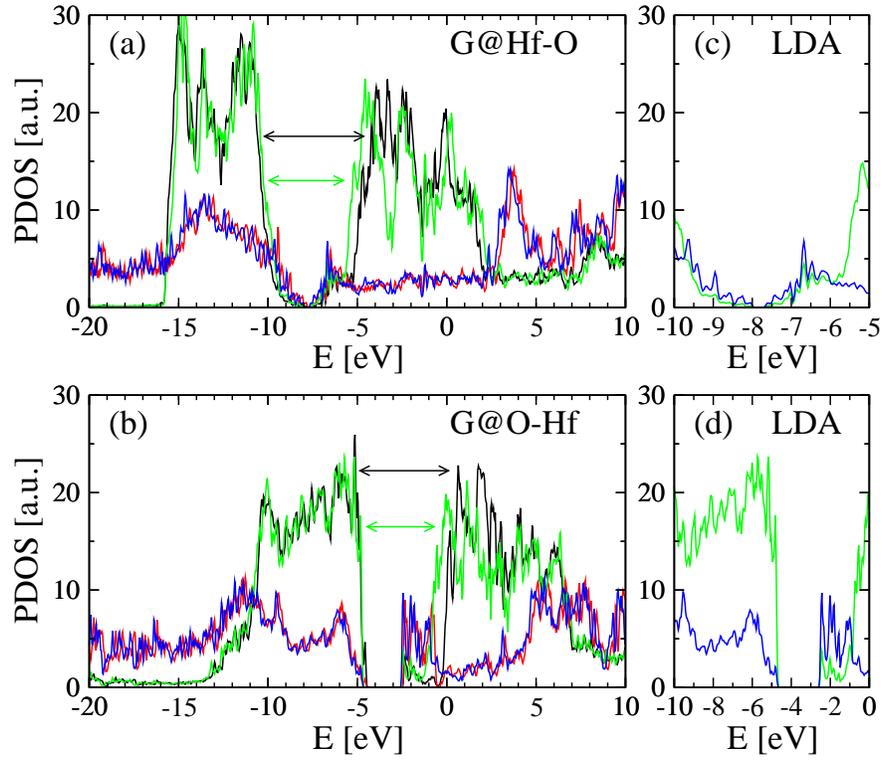} 
\caption{A comparison between LDA+U and LDA for (a) G@Hf-O and (b) G@O-Hf, with a detailed representation of the LDA PDOS indicated in (c) and (d) plots, respectively. Contributions corresponding to HfO$_2$ L5 interfacial layer and graphene are depicted: LDA+U -- HfO$_2$ (black), graphene (red); LDA -- HfO$_2$ (green), graphene (blue). Using LDA the bulk-like gap of the L5 layer is reduced with $\approx 1$ eV, but the contribution to the in-gap states is rather similar and consequently, also the gaps induced in graphene are about the same magnitude as for the LDA+U approach.}
\label{PDOS_LDAU_LDA}
\end{figure*}

\end{document}